\documentclass[twocolumn,oldversion,a4paper]{aa}
\usepackage{graphicx}
\usepackage{epsfig}
\usepackage{color}
\usepackage{txfonts}
\usepackage{natbib}
\bibpunct{(}{)}{;}{a}{}{,} % to follow the A&A style

\def\approxlt{\lower.2em\hbox{$\buildrel < \over \sim$}}
\def\approxgt{\lower.2em\hbox{$\buildrel > \over \sim$}}

\newcommand{\HI}{\mbox{H\,{\sc i}}}

\def\gtrsim{\mathrel{\hbox{\rlap{\hbox{\lower4pt\hbox{$\sim$}}}\hbox{$>$}}}}

\def\lesssim{\mathrel{\hbox{\rlap{\hbox{\lower4pt\hbox{$\sim$}}}\hbox{$<$}}}}

\def\la{\mathrel{\hbox{\rlap{\hbox{\lower4pt\hbox{$\sim$}}}\hbox{$<$}}}}
\def\ga{\mathrel{\hbox{\rlap{\hbox{\lower4pt\hbox{$\sim$}}}\hbox{$>$}}}}

\begin{document}

\authorrunning{Lehnert et al.}

\title{On the cosmic evolution of the specific star formation rate}

\titlerunning{Evolution of the specific star formation rate}

\author{M. D. Lehnert\inst{1}\thanks{email: lehnert@iap.fr},
W. van Driel\inst{2}, L. Le Tiran\inst{2,3},
P. Di Matteo\inst{2} \and M. Haywood\inst{2}}

\authorrunning{Lehnert et al.}

\institute{Institut d'Astrophysique de Paris, UMR 7095, CNRS,
l'Universit{\'e} Pierre et Marie Curie, 98 bis boulevard Arago, 75014
Paris, France
\and 
GEPI, Observatoire de Paris, UMR 8111, CNRS, Universit\'e Paris Diderot,
5 place Jules Janssen, 92190 Meudon, France
\and
Departamento de Astronomia, IAG/USP, Rua do Mat\~ao 1226, 05508-090,
Cidade Universit\'aria, S\~ao Paulo, SP, Brazil}

\date{Received/Accepted}

\abstract{The apparent correlation between the specific star formation
rate (sSFR) and total stellar mass (M$_{\star}$) of galaxies is
a fundamental relationship indicating how they formed their stellar
populations. To attempt to understand this relation, we hypothesize
that the relation and its evolution is regulated by the increase in
the stellar and gas mass surface density in galaxies with redshift,
which is itself governed by the angular momentum of the accreted gas,
the amount of available gas, and by self-regulation of star formation.

With our model, we can reproduce the specific SFR-M$_{\star}$ relations
at z$\sim$1--2 by assuming gas fractions and gas mass surface densities
similar to those observed for z=1--2 galaxies.  We further argue that
it is the increasing angular momentum with cosmic time that causes
a decrease in the surface density of accreted gas. The gas mass
surface densities in galaxies are controlled by the centrifugal support
(i.e., angular momentum), and the sSFR is predicted to increase as,
sSFR(z)=(1+z)$^3$/t$_{\rm H0}$, as observed (where t$_{\rm H0}$ is
the Hubble time and no free parameters are necessary). In addition,
the simple evolution for the star-formation intensity we propose is in
agreement with observations of Milky Way-like galaxies selected through
abundance matching.

At z$\ga$2, we argue that star formation is self-regulated by high
pressures generated by the intense star formation itself. The star
formation intensity must be high enough to either balance the hydrostatic
pressure (a rather extreme assumption) or to generate high turbulent
pressure in the molecular medium which maintains galaxies near the line
of instability (i.e. Toomre Q$\sim$1). We provide simple prescriptions
for understanding these self-regulation mechanisms based on solid
relationships verified through extensive study. In all cases, the most
important factor is the increase in stellar and gas mass surface density
with redshift, which allows distant galaxies to maintain high levels
of sSFR.  Without a strong feedback from massive stars, such galaxies
would likely reach very high sSFR levels, have high star formation
efficiencies, and because strong feedback drives outflows, ultimately
have an excess of stellar baryons.}


\maketitle

\section{Introduction}\label{sec:intro}

The evolution of the star formation rate (SFR) and the relation between
the specific star formation rate (SFR per unit stellar mass, sSFR) and
total stellar mass (M$_{\star}$) of galaxies has garnered considerable
observational and theoretical attention \citep[e.g.][]{elbaz07, daddi07,
elbaz11, weinmann11, stark13, behroozi13}. Observations of galaxies over
a wide range of redshifts suggest that the slope of the SFR-M$_{\star}$
relation is about unity \citep[e.g.][]{elbaz07,salmi12}, which implies
that their sSFR does not depend strongly on stellar mass. Specific
star formation rates increase out to z$\approx$2 \citep{elbaz07,daddi07,
daddi09, noeske07, dunne09, stark09, oliver10, rodighiero10, elbaz11}
and are constant, or perhaps slowly increasing, from z=2 out to
z=6, though with a large scatter, sSFR$\approx$2-10 Gyr$^{-1}$
\citep{feulner05,dunne09,magdis10b, stark13}.

It is important to emphasize that neither the observed SFR-M$_{\star}$
nor the sSFR-M$_{\star}$ relationship implies a correlation, but that
both are actually ridge lines in the distribution of actively star-forming galaxies
-- galaxies that are evolving passively or forming
stars at moderate rates lie below these relationships at a given
mass \citep[][]{rodighiero10,elbaz11, karim11}. Depending on epoch,
the fraction of passively evolving galaxies can be significant, in
particular among very massive objects \citep{karim11}.

Currently there is no widely accepted explanation as to why the
relative rate of growth of galaxies depends on redshift in this
manner \citep[e.g.][]{dutton10,khochfar11,weinmann11} other than it is
likely to be a complex interaction between the gas supply, the rate at
which gas is transformed into stars and material lost from the galaxy
(and halo) through outflows \citep[e.g.][]{bouche10, dave11, shi11,
lilly13}. Theoretically, the rate of cosmological baryonic accretion onto
a galaxy halo is expected to be a function of mass and time, depending
on redshift as $\dot{\rm M}_{\rm acc}$/M$\propto$(1+z)$^{2.25-2.5}$
\citep{neistein08, dekel09, dekel13}, contrary to the observed
relationship sSFR $\propto$ (1+z)$^3$ at z$\la$2 \citep{oliver10,
elbaz11}, while at higher redshifts it either remains constant or
increases more slowly \citep[e.g.][]{stark13}. Of course, there
are several caveats in making a direct link between the specific halo
accretion rate and the sSFR, such as assuming that the ratio of halo
mass to stellar mass is constant at constant halo mass with redshift
\citep{behroozi13}. While many variables come into play in determining
the mass accretion rate, it appears that the general increase in the
sSFR with redshift is not simply controlled by the gas supply, and that
other processes must come into play.

Any plausible explanation must reconcile the available gas supply with
the evolution of the sSFR. Currently, the most direct ways of relating
the specific growth of galaxies to the specific accretion rate is to
use AGN and starburst driven outflows and gas consumption timescales to
regulate the star formation in galaxies \citep[e.g.][]{dekel09,lilly13}.
The effects of feedback from massive stars and active galactic nuclei
allow this direct coupling of star formation with the gas supply to be
broken \citep[e.g.][]{peirani12,L13}. This decoupling is important as
not only do we need the relative growth of galaxies not to track the gas
supply too closely, as observed in the evolution of the sSFR, but also
because
baryonic mass fraction in galaxies is small and does not follow the halo
mass function \citep{baldry08, papastergis12} suggesting that either a
fraction of the baryons are not accreted or they are efficiently removed
from the galaxy. Galaxy growth and baryon content must be limited by the
way in which gas is accreted, cools and collapses, or alternatively,
by processes that are internal to the galaxy or the physics of star
formation \citep[e.g.][]{dutton10}. In fact, breaking this coupling
may be necessary to explain some aspects of the evolution of the sSFR
within the context of simulations, which often produce too little star
formation at recent epochs and exhibit a positive correlation between
the sSFR and stellar mass \citep[similar to that of the specific dark
matter accretion rate;][]{weinmann13}.

The lack of a direct coupling between accretion and star formation would
favor an explanation of the sSFR-M$_{\star}$ relationship through local
processes such as star formation controlling the pressure of the ISM,
and hence self regulation \citep[e.g.][]{silk97}. One observational
signature of this self regulation is galaxy-wide outflows, which are
observed in intensely star-forming galaxies across all cosmic epochs
\citep[e.g.][]{lehnert96,weiner09,steidel10}. This complex gas physics
and the strong interaction between phases of the interstellar (ISM),
inter-halo and intra-halo medium are processes which are difficult to
simulate currently because of both a lack of computational power and our
general ignorance of what processes to model, and how \cite[e.g.][
and references therein]{silk12}.

Because there are many gaps in our understanding of how stars form and
the physics of the ISM, it is difficult to model the processes that
may lead to the SFR-M$_{\star}$ relation and the evolution of the sSFR
from first principles, and we thus need to take another approach. In
this paper, we use simple analytical arguments to demonstrate that both
relationships arise naturally in a scenario where star formation is not
only limited by gas supply, but also by self-regulation during phases
of rapid galaxy growth, when the supply of accreting gas is large. When
the gas supply drops below the rates necessary to maintain the high
rates of star formation required for self regulation, secular processes
become important, including the self-gravity provided by stellar disks
\citep[e.g.][]{shi11}. Motivated by the observed slope of the sSFR-z
relationship at z$\la$2, we suggest that the decline in sSFR is not
only driven by declining gas fractions, but also by an evolution in the
centrifugal support of the accreting gas (i.e., angular momentum), and
by gas and stellar mass surface densities through a generalized Schmidt
law \citep{dopita94, shi11}. Thus, the sSFR-z relationship may suggest
two epochs of galaxy growth, first of self regulation at z$\ga$2, which
limits the ensemble specific star formation rate, followed by an epoch of
secular growth at z$\la$2, where galaxies are prevented from consuming
their gas too quickly and efficiently because the gas is accreted with
relatively high angular momentum.

The paper is organized as follows: In \S~\ref{sec:SFRmass}, we present a
simple model to explain the detailed evolution of the sSFR from z=0 to
2 within the contex of the ISM pressure and a generalized Schmidt law.
We find that to get good agreement with the model and the data, the
required gas mass surface densities and gas fractions are consistent
with those that have been observed for local and distant galaxies. In
\S~\ref{sec:evol}, we discuss the evolution of the sSFR in a more
general context, specifically commenting on why there is a change in
the apparent evolution of the sSFR above and below z$\approx$2. Finally,
in \S~\ref{summary}, we provide a brief summary.

\section{The ridge line in the SFR-M$_{\star}$
plane}\label{sec:SFRmass} % 2

The apparent SFR-M$_{\star}$ relation and its evolution from
z$\approx$0--7 \citep[e.g.][]{elbaz07,daddi07,stark09, oliver10,
magdis10b, stark13} provides valuable insight into how galaxies
convert their gas into stars. As a ridge line to the distribution of
galaxies in the SFR-M$_{\star}$ plane it shows in particular how the
sSFR of vigorously star-forming galaxies evolves with stellar mass and
redshift. Although the slope of the ridge in this plane is roughly the
same at every epoch, high-redshift galaxies can reach much higher sSFR
values than galaxies at more moderate redshifts (z$\la$2).

\subsection{A simple model relating overall SFR to ISM
pressure}\label{subsec:simplemodel}

We will now show that the SFR-M$_{\star}$ relationship may be explicable
through a simple model which relates the overall star formation
rate in galaxies to the overall pressure of their ISM \citep{silk97,
blitz06,silk09, shi11}. Here, the star formation rate is limited to
regimes where the pressure induced by mechanical energy injection of star
formation remains below the hydrostatic mid-plane pressure in galaxies
\citep[e.g.][]{lehnert96}.

One simple analytic way of investigating if pressure is indeed the
driver of the star-formation intensity is by relating the star-formation
intensity to gas and total mass surface densities through a generalized
Schmidt law, as $\Sigma_{\rm SFR}$=$\epsilon_{\rm GS}$$\Sigma_{\rm
gas}^{3/2}$$\Sigma_{\rm total}^{1/2}$, where $\Sigma_{\rm gas}$ is
the gas mass surface density and $\Sigma_{\rm total}$=$\Sigma_{\rm
gas}$+$\Sigma_{\star}$, where $\Sigma_{\star}$ is the stellar mass surface
density \citep{dopita94,silk09}, and $\epsilon_{\rm GS}$ is an efficiency
factor. The units of $\epsilon_{\rm GS}$ are those of the gravitational
constant, G, divided by a velocity (pc$^2$ M$_{\sun}^{-1}$ yr$^{-1}$). It
is not clear if $\epsilon_{\rm GS}$ is constant as a function of redshift
or galaxy mass \citep[see][]{dopita94}. For now, we will assume it to
be constant, and argue this case later in \S~\ref{constantefficiency}.

The pressure in the ISM can be related to gravity or turbulence through
${\rm P}_{\rm gas}$=$\rho_{\rm gas}\sigma_{\rm gas}^2$=$\frac{\pi}{2}
{\rm G}\Sigma_{\rm gas}\Sigma_{\rm total}$, where ${\rm P}_{\rm gas}$
and $\rho_{\rm gas}$ are the gas pressure and density, respectively. $G$
is the gravitational constant, and $\sigma_{\rm gas}$ is the velocity
dispersion of the turbulent gas.

Combining these implies that $\Sigma_{\rm SFR}\propto {\rm
P_{gas}}(\Sigma_{\rm gas}/\Sigma_{\rm total})^{1/2}$. By doing this
we are simply emphasizing the role played by interstellar pressure in
regulating the star formation in galaxies. If supernovae are driving the
turbulence in the ISM, then star formation would be self-regulating
in such a scheme. This formulation includes hydrostatic as well
as turbulent pressure driven by star formation. For simplicity, we
adopt a simple relation for the hydrostatic pressure and not the more
general relation which takes into account the possibility of different
dispersions in the gas and stars \citep{elmegreen89, elmegreen93}. We
will discuss the impact of this choice at the end of this section (see
also \S~\ref{subsubsec:Q1}).

Returning to the generalized Schmidt law given above, we can
estimate $\epsilon_{\rm GS}$ using the argumentation and results from
\citet{silk09} (their equation 4). Adopting reasonable parameters for the
fraction of gas in molecular clouds, the covering fraction of dense gas,
the momentum relative to the energy of supernovae and the ratio of the
ISM pressure to molecular cloud pressures \citep[see Appendix A and
][for details]{silk09}, and that the gas and stars cover the same extent,
we can relate the star formation rate and the gas surface density, as

\begin{equation}
{\rm SFR} = 6.5\times10^{-12} {\rm M}_{\star}\ \frac{{\rm f}_g^{1/2}}{(1-{\rm f}_g)} \Sigma_{\rm gas} {\rm M_{\sun}} \ {\rm yr^{-1}}
\label{eqn:SFR}
\end{equation}

where f$_g$ is the molecular gas fraction, $\Sigma_{\rm
gas}$/$\Sigma_{\rm total}$.

Does this formulation of the SFR agree well with observations?
Unfortunately, many galaxies for which the necessary data are available
(such as the Milky Way) lie well below the upper envelope of the
SFR-M$_{\star}$ plane \citep {elbaz11,leroy08}, which could bias
the resulting gas fraction and gas and stellar mass-surface densities
necessary to explain the relations between sSFR or SFR and M$_{\star}$.
At M$_{\star}$$\sim$10$^{10.5}$ M$_{\sun}$, the local relation has
an SFR$\approx$1.5 M$_{\sun}$ yr$^{-1}$ \citep[e.g.][]{elbaz11}.
Average values for the gas fraction and gas mass surface densities for
statistical samples of nearby late-type (i.e. star forming) galaxies
are approximately $<$$\Sigma_{\rm gas}$$>$$\sim$20 M$_{\sun}$ pc$^{-2}$
and f$_{g}$$\sim$10\% \citep{young95, young89, bigiel12}. These values
are consistent with the observed star formation rates giving us some
confidence that this approach is plausible.

Only at z$\la$2 do we have both well constrained SFR-M$_{\star}$
relationships and a reasonable number of estimates of the
molecular gas content of galaxies through CO observations
\citep[e.g.][]{elbaz07,daddi07}, estimated at f$_{g}$$\sim$0.2-0.5
and $\Sigma_{\rm gas}$$\sim$100-1000 M$_{\sun}$ pc$^{-2}$
\citep[][]{daddi10,aravena10,dannerbauer09,tacconi10,tacconi13}.
$\Sigma_{\rm gas}$=150 and 330 M$_{\sun}$ pc$^{-2}$, and f$_{g}$=0.25
and 0.45 at z=1 and 2 respectively, yields relationships that are
consistent with the best fits to the ridge line in the SFR-M$_{\star}$
plane \citep[Fig.~\ref{fig:elbaz};][]{elbaz07,daddi07}. Within the
context of this model, the scatter in the data about the mean relations is
due to the variation in the gas mass surface densities and gas fractions
which is consistent with observations \citep{tacconi13} and that there
is little or no mass dependence on the sSFR \citep[see][]{abramson14}.

\subsection{Choices made for this model}\label{subsec:choices}

We made several choices to conduct this analysis which warrant further
discussion.

\subsubsection{Exponents in the generalized Schmidt
law, m+n=2}\label{subsubsec:exponents}

The first is our choice of the specific exponents and their sum in the
generalized Schmidt law. A law of this form can be justified either from
theoretical or observational arguments.

Theoretically, a generalized Schmidt law can be justified through a
cloud-cloud collision model in a turbulent ISM where the collision rate
is determined by the stellar energy injection rate into the ISM, and
where the clouds are confined by the ambient ISM \citep[e.g.][]{silk09,
inoue13}. Turbulence may be driven by the energy injection from young
stars at high SF intensities \citep[e.g.][]{agertz09}.

Observationally, interpreting the relationship between stellar mass
surface density and star formation intensity in local disk galaxies can
lead to a star formation rate formula that depends on both gas and stellar
mass surface densities, such as the generalized Schmidt law of the form,
$\Sigma_{\rm SFR}\propto\Sigma_{\rm gas}^{m}\Sigma_{\rm total}^{n}$, with
m+n$\approx$2, found by \citet{dopita94} \citep[see also ][]{shi11}. While
formally, based on theoretical arguments, \citet{dopita94} favored n=1/3
and m=5/3, as they pointed out, changing an assumption in their analysis
would push n to 1/2 and likely decrease m. Indeed, within the context
of our analysis, as long as n+m=2, the impact of the precise choice
of m and n only has an impact on the dependence of the SFR on the gas
fraction in the form of the function.  Therefore, for the purpose of this
discussion, it is sufficient to say that the analysis of \citet{dopita94}
is consistent with our assumption of n=1/2,m=3/2, and that changing the
exponents (as long as m+n=2) will make little difference overall.

%fig 1
\begin{figure}
\includegraphics[width=8.5cm]{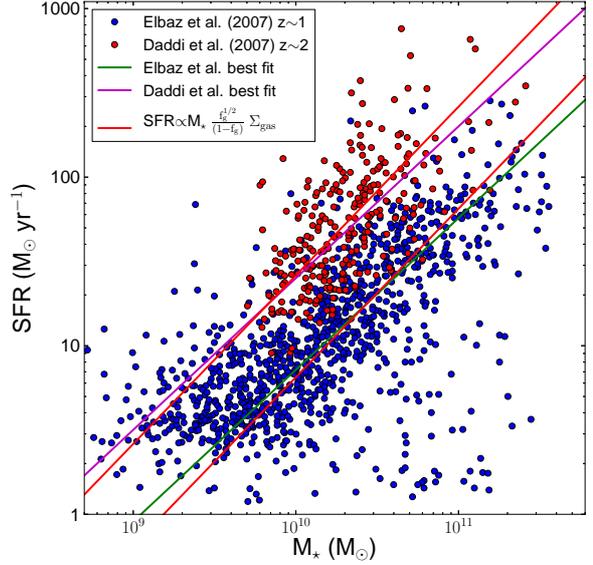}
\caption{Star formation rate (SFR, in M$_{\sun}$ yr$^{-1}$) as function of total stellar mass (M$_{\star}$, in M$_{\sun}$) in two galaxy samples. The data points are estimates of the star formation rate and stellar mass
for samples of galaxies at z$\sim$1 \citep[blue circles;][]{elbaz07} and
z$\sim$2 \citep[red circles;][]{daddi07}. The green and purple solid lines indicate the best fits to the data sets at z$\sim$1 and z$\sim$2 from the original papers. The two red lines indicate our simple star formation model with the best scaling parameters adopted (see text for details).}
\label{fig:elbaz} 
\end{figure}

\subsubsection{Exponents in the generalized Schmidt law,
m+n$\la$2}\label{subsubsec:exponents2}

While some studies have shown that the slope of the ridge line in
the SFR-M$_{\star}$ plane is about one \citep[e.g.][]{salmi12}, other
studies suggest it is less than unity \citep[e.g.][]{rodighiero10}.
Recently, \citet{abramson14} have suggested that slopes less
than one for the local ridge line may be due to the contribution from
the bulge. If we relax the requirement that m+n=2, which we adopted
because of theoretical arguments \citep{silk09} and its consistency with
observations \citep[][]{dopita94, shi11}, then we can accommodate slopes
less than unity for the SFR-M$_{\star}$ relation (and negative slopes
for the sSFR-M$_{\star}$ relation).

Suppose we consider a generalized Schmidt law of the form, $\Sigma_{\rm
SFR}\propto\Sigma_{\rm gas}^{1.3}\Sigma_{\rm total}^{0.5}$ and require
that the total star formation rate should be proportional to the gas mass
surface density, for consistency with the pressure arguments we made in
\S~\ref{subsec:simplemodel}. Reformulating this leads to ${\rm SFR}\propto
{\rm M}_{\star}^{0.8}\ \frac{{\rm f}_g^{0.3}}{(1-{\rm f}_g)^{0.8}}
\Sigma_{\rm gas}$. Generalized Schmidt laws can be constructed which can
accommodate different slopes. However, a full discussion of this point and
its theoretical justification is beyond the scope of the current paper.

\subsubsection{Equal velocity dispersions of gas and
stars}\label{subsubsec:dispersions}

Another choice we have made in this model is that the velocity dispersion
of gas and stars are roughly equal in the equation of hydrostatic
pressure.

From fitting the hydrostatic pressure in a self-gravitating plane,
\citet{elmegreen93} suggested the more appropriate relation is,
${\rm P}_{\rm gas}$=$\frac{\pi}{2}{\rm G}\Sigma_{\rm gas}(\Sigma_{\rm
gas} + (\sigma_{\rm gas}/\sigma_{\rm stars})\Sigma_{\rm stars}$).
For high redshift galaxies, there are few robust estimates of the
stellar velocity dispersion, but various estimates suggest that
their gas and stars have comparable velocity dispersions. Certainly,
the velocity dispersions of both the warm ionized gas and perhaps the
molecular gas are high at z$\sim$1-3 \citep[$\sim$30-200 km s$^{-1}$;
e.g.][]{L09,law09,L13,swinbank11,tacconi13}.

In local late-type spiral star forming galaxies, a significant
fraction of the gas mass may reside in the molecular phase, in which
typically $\sigma_{\rm gas}$ is a few to 10 km s$^{-1}$, while for the
neutral Hydrogen, \HI, it is somewhat higher and may be driven by the
intensity of star formation \citep{tamburro09, wilson11}. The bulk of
the stellar mass has an even higher dispersion, $\sim$20-100 km s$^{-1}$
\citep[e.g.][]{bottema93,neistein99}. Although for local disk galaxies,
the $\sigma_{\rm gas}/\sigma_{\rm stars}$ ratio is relatively small,
it is also likely that stellar disks are heated during their evolution
\citep[e.g.][]{qu11, masset97} and that at an early evolutionary phase
their stellar velocity dispersions were much higher \citep{bovy12,
haywood13, bird13, brook12, L14}.

We can make a rough estimate of the stellar velocity dispersions in
distant galaxy disks are about 90 km s$^{-1}$, through the relation,
H=$\sigma^2$/($\pi$ G $\Sigma_{\rm total}$), where H is the disk height,
$\sigma$ is the velocity dispersion and $\Sigma_{\rm total}$ is the
disk mass surface density. To make this estimate, we adopted H$_{\rm
z\sim2}$=1 kpc \citep{elmegreen06} and $\Sigma_{\rm total,z\sim2}$
= 350 M$_{\sun}$ pc$^{-2}$ \citep{tacconi10, fs11, mosleh12}.
To check for consistent and comparison, for the Milky way, we adopt
$\sigma_{\rm stars, MW}$ = 25 km s$^{-1}$ for the young thin disk stars
\citep[e.g.][]{bond10}, $\Sigma_{\rm total, MW}$= 70 M$_{\sun}$ pc$^{-2}$
\citep[e.g.][]{zhang13}, and a thin disk height, H$_{\rm MW}$=350 pc
\citep[e.g.][]{juric08}.

This value of the velocity dispersion is close to those inferred for the
gas in z$\sim$2 galaxies \citep[][]{L09, law09, swinbank11, L13}. Without
stronger constraints, it seems reasonable to assume that $\sigma_{\rm
gas}/\sigma_{\rm stars}$$\approx$1 in high redshift galaxies. However, if
the gas out of which stars are forming were to have a dispersion smaller
than that of the previous generations of stars, like in local galaxies,
this would only make a relatively minor difference within the context
of our scenario -- in the sense that this would require somewhat higher
gas surface densities to explain the location of the SFR-M$_{\star}$
ridge line.

\subsubsection{A constant efficiency factor in the generalized Schmidt
law}\label{constantefficiency}

Finally, and perhaps most importantly, we assumed that the efficiency
factor of the generalized Schmidt law, which is not unit-less, is
constant.

\citet{dopita94} derived a Schmidt law with n=1/3 and m=5/3, adopting
a star formation self-regulation model where the efficiency factor
depends on the inverse of escape velocity (rotation speed), weakly on
the sum of the gas and stellar disk heights, and on a constant roughly
proportional to the star formation efficiency. Their analysis implies
that the efficiency factor is not constant as we assumed.

Taking a somewhat different approach than that of \citet{dopita94}, we
can derive a Schmidt law with n=1/2 and m=3/2, for which we will argue
that a constant efficiency factor is a reasonable assumption. Adopting
a formalism of \citet{dopita94} and \citet{silk09}, we start by assuming
that the star formation intensity is proportional to the gas mass surface
density and the inverse cloud-cloud collision timescale, $\Sigma_{\rm
SFR}$=$\beta\Sigma_{\rm gas}/{\rm t}_{\rm c-cc}$, where ${\rm t}_{\rm
c-cc}$ is the cloud-cloud collision timescale, for which \citet{silk09}
derived the relation, ${\rm t}_{\rm c-cc}$=f$_{\rm cl}^{-1}$(P$_{\rm
cl}$/P$_{\rm gas}$)$^{1/2}$($\Sigma_{\rm tot}$/$\Sigma_{\rm
gas}$)$^{1/2}$(H$_{\rm gas}$/$\sigma_{\rm gas}$), where f$_{\rm cl}$ is
the mass fraction of gas in clouds, P$_{\rm cl}$ is the pressure in the
clouds, P$_{\rm gas}$ is the ambient average gas pressure, and H$_{\rm
gas}$ is the gas disk height. Substituting this into the equation for
$\Sigma_{\rm SFR}$ and using the relation between $\Sigma_{\rm tot}$
and $\sigma_{\rm gas}$, as well as the gas pressure due to turbulent
motions (as discussed at the beginning of this section), yields,
$\Sigma_{\rm SFR}$= $\gamma$ f$_{\rm cl}$ ($\rho_{\rm gas}$/P$_{\rm
cl}$)$^{1/2}$$\Sigma_{\rm gas}^{3/2}$$\Sigma_{\rm total}^{1/2}$, where
$\gamma$ is a constant of proportionality which can be interpreted as a
star-formation efficiency. \citet{silk09} make a very similar derivation, also leading to a Schmidt law with exponents n=1/2 and m=3/2.

This relation is consistent with the global ``Schmidt-Kennicutt''
relation, $\Sigma_{\rm SFR}$=C$_{\rm S-K}$$\Sigma_{\rm gas}^{3/2}$,
where C$_{\rm S-K}$ is a constant of proportionality and $\Sigma_{\rm
total}^{1/2}$ has been subsumed into C$_{\rm S-K}$. At z$\la$1, the
stellar mass surface densities of disk galaxies are roughly constant
\citep[Freeman's law; ][]{barden05} and the total mass surface densities
are dominated by the stars, not the gas \citep{young95, young89}. In
principle, the Schmidt-Kennicutt relation could be made to evolve
due to increasing gas fraction, increasing star formation rates as a
function of mass and increasing star formation efficiency with increasing
redshift. However, no significant evolution is seen in the relation out
to z$\approx$2 despite the changes in the ensemble of the population of
star forming galaxies over this cosmic epoch \citep{bouche07}. This,
together with the approximate constancy of the stellar mass surface
density, suggests empirically that the scaling of the Schmidt law is
approximately constant as a function of redshift, or at most changes
within the intrinsic scatter of the relation. This argues that the scaling
relation in our generalized Schmidt law likewise does not evolve strongly
with redshift (at least out to z$\approx$2).

In addition, \citet{shi11} performed a similar analysis to the one
presented here for an extended Schmidt law with different exponents,
$\Sigma_{\rm SFR}$$\propto$$\Sigma_{\rm gas}^{k}$$\Sigma_{\star}^{l}$,
which also emphasizes the important role played by stellar mass
surface density in regulating the SFR. To determine the
exponents of the law, they fit the data for a small sample of local
and distant (z$\approx$2) galaxies. Fitting their data set with our
generalized Schmidt law, we find that the fit is significant and has
a scatter comparable to their best fits of the extended Schmidt law or
a Schmidt-Kennicutt relation. Since, in this analysis, we are fitting
galaxies over a wide range of star formation intensities and redshifts,
this suggests that the scaling of the generalized Schmidt law does not
evolve strongly with redshift and that there is a roughly constant
proportionality to the star formation intensity and $\Sigma_{\rm
gas}^{3/2}\Sigma_{\rm total}^{1/2}$.

\section{Cosmic evolution of the sSFR}\label{sec:evol}

\begin{figure}
\includegraphics[width=8.95cm]{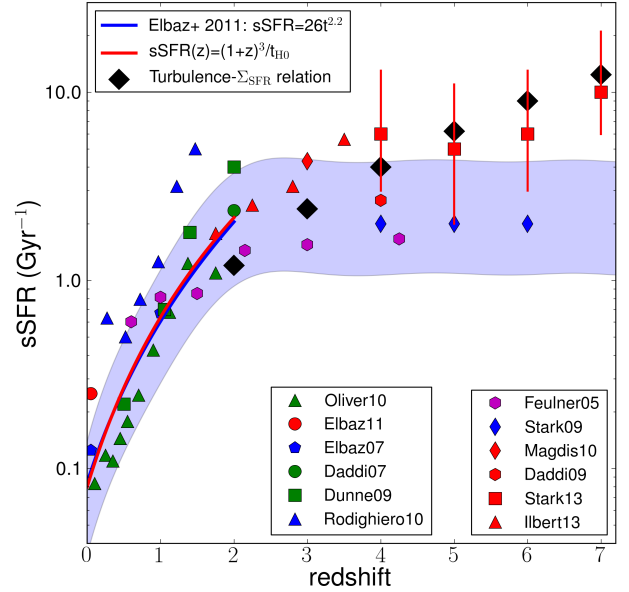}
\caption{The specific star formation rate (sSFR, in Gyr$^{-1}$) as a
function of redshift. The various points represent measurements from the
literature at M$_{\star}$$\sim$10$^{10}$ M$_{\sun}$; see the references in
the legend at the bottom right.  We note that we have specifically
shown the determinations from \citep{stark13} with their uncertainties
(red squares) because there is some controversy as to whether or not there
is any evolution in the sSFR with increasing redshift beyond z=2. Since
the slope of the sSFR-M$_{\star}$ relation is approximately zero, the
rate at which the sSFR evolves is largely independent of M$_{\star}$,
except at the highest masses. The lines represent the best-fit relation
from \citet{elbaz11} over the redshift range 0 to 2 (blue line) and a
simple scenario, sSFR (z) = (1+z)$^3$/t$_{\rm H0}$ (where t$_{\rm H0}$ is
the Hubble time at z=0, red line), and the blue squares are the estimates
for our hypothesis relating the turbulence to the star formation intensity
(see text for details). The blue shaded region represents the scatter in
the observed sSFR values ($\pm$0.3 dex). This rendition of the evolution
of the sSFR is inspired by a similar plot in \citet{weinmann11}.}
\label{fig:sSFRevol} \end{figure}

The zero-point of the sSFR-M$_{\star}$ relation at z$\la$2
is observed to evolve as sSFR=26t$^{-2.2}$$\propto$(1+z)$^{3}$
\citep{elbaz11,oliver10}. At z$\ga$2, the sSFR-M$_{\star}$ relation
perhaps reaches a plateau or increases only slowly with redshift
(Fig.~\ref{fig:sSFRevol}). Although the scatter in the sSFR evolution
plot is large, especially at the highest redshifts, the slowing of its
increase beyond redshifts of about 2 appears to be robust.

The evolution of the sSFR through cosmic time does not appear to
track the expected accretion rate of matter into the halos of galaxies
\citep[e.g.][]{weinmann11}. Models where the gas supply directly drives
the growth of the stellar and gaseous masses of galaxies have a number
of difficulties compared to the results from observations. For example
if the gas supply and galaxy growth rate occurred in lock step, the mass
dependence of the sSFR would have a positive slope, whereas observations
show no or a slightly negative slope \citep{abramson14}, and the
sSFR would be much higher than is observed in the early universe and
lower than observed in the local universe \citep{silk12}.

These contradictions suggest that there are regulatory processes that
control the baryon content and its distribution as gas is accreted onto
a galaxy, that star formation must be kept relatively inefficient, that
much of the accreted gas must ultimately be ejected and/or accreted
into the halo with long cooling times, and in the local universe, that
a sufficient gas supply must be maintained in galaxies to support the
average sSFR, which is above the specific mass accretion rates estimated
from the cosmic web.

A regulatory process (or processes) which limits the ultimate sSFR a
galaxy can reach would also naturally explain the high star formation
intensities that galaxies can support with increasing redshift and the
fact that the most intensely star forming galaxies appear to lie above
the ridge line of the main sequence \citep{wuyts11}.

We will now explore several processes which might limit or regulate the
star formation intensities of galaxies and hence control the evolution
of the sSFR of the ensemble of galaxies with epoch -- i.e., the angular
momentum content of the gas as it is accreted \citep{danovich14},
and feedback and self-regulation by intense star formation.

\subsection{The evolution of angular momentum}\label{angmoevolution}

To explain the evolution of the sSFR with redshift, we suggest
that a crucial role is played by the relative amount of angular
momentum that the gas acquires and is able to maintain during its
accretion onto a galaxy. The specific accretion rate, i.e., the
accretion rate per unit stellar mass, is generally higher than the
specific star formation rate of galaxies at high redshift, z$\ga$2
\citep[e.g.][]{dekel09,weinmann11,dave11}. Galaxies at z$\ga$2 show
significant evolution in their half-light radii, which decreases as
(1+z)$^{-1.2\pm0.1}$ in the stellar mass range 9 $\la$ log M$_{\star}$
(M$_{\sun}$) $\la$10.5 \citep{oesch10, mosleh12}, which means that
galaxies at higher redshifts have higher stellar mass surface
densities and, by corollary, higher gas mass surface densities
\citep[e.g.][]{tacconi13}. This increase in surface density with
redshift implies that as the gas accretion rate increases with redshift
\citep{dekel09}, the angular momentum of accreted gas is overall lower,
allowing it to collapse to higher mass surface densities.

%fig 2

Whatever the exact cause of the relatively low angular momentum of
the accreting gas in high redshift  galaxies, which  allows them to reach
high surface densities \citep{danovich14}, the high gas  surface densities
enables high star formation intensities through the relationship between
star formation intensity and gas surface density, which has an exponent of
approximately unity \citep[the Schmidt-Kennicutt relation; ][]{leroy13}.
High intensity star formation is precisely the regime where the effects
of stellar feedback are likely to play an important role in preventing
stars from forming efficiently and to limit the final baryon mass fraction
of galaxies and thus to inhibit galaxies from growing in lock step with
the gas supply, as observed. To keep the total baryon fractions low,
these feedback effects must also include efficient outflows.

At lower redshifts (z$\la$2), the gas is likely accreted onto the
galaxy proper with higher total and specific angular momentum
than at higher redshifts, which may be due to the formation of gas
streams between asymmetric voids, the larger virial radii of halos at
lower redshifts \citep{pichon11, codis12} and the central mass surface
density of late-type galaxies at z$\la$1 apparently already having been
{\it mis en place} \citep{barden05} so much of the gas is accreted onto
the outskirts of the disk.

The difference in the sSFR evolution below and above z$\sim$2 suggests
that at redshifts above $\sim$2, the observed increasing stellar mass
surface densities would necessitate a different regulatory mechanism due
to changes in the angular momentum of the gas as it is accreted into
the halo. How much angular momentum the gas has and/or retains or gains
as it falls into the halo and accretes onto the galaxy as a function
of redshift \citep[][]{pichon11, ceverino12, dubois13} is therefore
perhaps the most important factor in determining what processes may affect the
evolution of the sSFR in relation to the specific gas accretion rate,
as we will now explore.

%fig 3
\begin{figure}
\includegraphics[width=8.95cm]{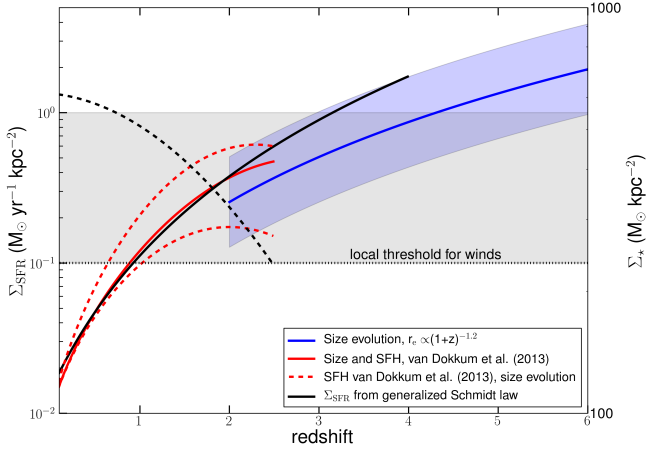}
\caption{The star formation intensity ($\Sigma_{\rm SFR}$,
in M$_{\sun}$ yr$^{-1}$ kpc$^{-2}$) as a function of redshift. The
evolution of the generalized Schmidt law (black line) compared to the
evolution of abudance-matched star-forming galaxies (i.e., galaxies
with co-moving number densities like that of MW mass galaxies in the
local Universe) for several types of size evolution \citep[r$_{\rm
e}$=3.6(M$_{\star}$/5.0$\times$10$^{10}$)$^{0.27}$ -- solid red line;
r$_{\rm e}$=3.6(1.+z)$^{-0.55}$ -- upper red dashed line; r$_{\rm e}$=3.6
kpc -- lower dashed red line; see][ for details]{vandokkum13}. We also
show the evolution of the star formation intensity for high redshift
galaxies with an sSFR = 2 Gyr$^{-1}$, a stellar mass of 5$\times$10$^{9}$
M$_{\sun}$, and a size evolution, r$_{\rm e}$=2.5((1+z)/3)$^{-1.2}$ kpc
\citep[blue line and shaded region representing a scatter of $\pm$0.3
dex; see][]{oesch10, mosleh12}. The evolution of the stellar mass surface
density, $\Sigma_{\star}$ (M$_{\sun}$ kpc$^{-2}$; right hand ordinate) of
MW-like galaxies (black dashed line) is a factor of about 3 but includes
the growth of the bulge as well as the disk \citep{vandokkum13}. We also
indicate (dotted line) the star formation intensity threshold for local
starburst galaxies to drive winds and indicated the possible range of
thresholds from local and distant galaxies \citep{lehnert96, heckman03,
LT11b, newman12a}.}
\label{fig:SFIvsz}
\end{figure}

\subsection{The declining sSFR at z$\la$2}\label{declineinsSFR}

Why does the sSFR decline with decreasing redshift below z$\approx$2? The
gas supply from cosmological gas accretion appears to increase with
redshift as $\dot{\rm M}_{\rm acc}$/M$\propto$(1+z)$^{2.25-2.5}$
\citep{dekel09, dekel13} and at z$\la$2 it is predicted to
be below the level necessary to support the sSFR observed in the ensemble
of galaxies \citep{dekel09, weinmann11, dave11}. Perhaps the decline is
regulated by the angular momentum of the accreted gas and that returned
by the stellar population as it ages.

The radius of galaxies supported by their own centrifugal force
(angular momentum) is expected to evolve as $\approx$(1+z)$^{-1.5}$
\citep[e.g.][noting that at low redshifts, below about
z$\approx$0.3-0.5, the slope as a function of redshift is
shallower]{mo98}. For galaxies of the same total mass over all epochs
\citep[Fig.~\ref{fig:sSFRevol};][]{weinmann11}, simple geometric
scaling implies, $\Sigma_{\rm total}$=M$_{\rm total}$/2$\pi$r$_{\rm
e}^{2}$, and if gas in the galaxies is centrifugally supported then
$\Sigma_{\rm total}$ evolves as $\approx$(1+z)$^{3}$. Observations
show that the stellar radius of galaxies evolves somewhat more slowly
than this at z$\ga$1-2 \citep[(1+z)$^{-1.2\pm0.1}$, e.g.][]{mosleh11,
mosleh12}. In contrast, \citet{barden05} find little evidence for
a significant ($\ga$30\%) increase in the sizes and stellar mass
surface densities of spiral galaxies from z=1 to 0. The declining gas
mass surface densities observed from z=2 to z=0 are consistent with
evolving to a rate proportional to (1+z)$^{3}$ (\S~\ref{sec:SFRmass}
and Fig.~\ref{fig:elbaz}).

The gas supply from accretion is declining and being incorporated into
the galaxy with high angular momentum, implying it likely ends up mostly
in the outer disk regions \citep{stewart11}. This is consistent with the
observation that the \HI\ mass surface density does not depend strongly
on radius in individual galaxies \citep{bigiel12}. Thus increasing the
angular momentum of the gas as the redshift decreases is a robust way
of growing galaxies in a ``centrifugally supported'' way as suggested
in the models of, e.g., \citet{mo98}. We note however that the relation
which best matches the evolution of the sSFR
out to $\sim$2 is neither arbitrary nor actually a fit to the data. We
would expect the stellar mass surface densities to scale as the Hubble
time, t$_{\rm H0}$ \citep{mo98} and in fact, that is what sets the
zero point for the relation shown in Fig.~\ref{fig:sSFRevol}, namely,
sSFR(z)=(1+z)$^{3}$/t$_{\rm H0}$, a relation indistinguishable from that
of \citet{elbaz11}. This scaling comes from cosmological considerations,
namely, the change of the structure of dark matter halos (increasing
virial radii, masses, etc.) and the overall growth of galaxies. Thus
to explain the observed relation requires no free parameters. Moreover,
within the context of gas accretion and mass return, one could argue that
the scaling we adopted for the generalized Schmidt law analysis is also
constrained within the context of our simple hypothesis, although based on
parameters estimated for the ISM of galaxies and through comparison with
measurements of the sSFR, gas mass surface densities and gas fractions.

We can extend this analysis further. Equation~\ref{eqn:SFR} is also a simple
relation between the star formation intensity and the stellar and gas
mass surface densities, which was obtained by multiplying both sides of
the equation by the disk surface area to yield a relation
between the SFR and M$_{\star}$; see Appendix A). Since sSFR(z) evolves
as (1+z)$^{3}$, this implies that $\Sigma_{\rm SFR}$ would also evolve
as (1+z)$^{3}$ in the context of our model. An evolution of the star
formation intensity as (1+z)$^{3}$ is shown in Fig.~\ref{fig:SFIvsz},
whose zero-point was chosen based on the relation between the disk
scale length of local disk galaxies \citep{fathi10} and star formation
rates of galaxies with stellar masses similar to the MW \citep{elbaz07,
lara-lopez13}. We compare this to the evolution a function of redshift
of the star formation intensity of MW-like galaxies selected through
abundance matching by \citep{vandokkum13}, who characterized the size
evolution of these MW-like galaxies as a function of redshift and
of stellar mass, as shown in Fig.~\ref{fig:SFIvsz} together with,
for completeness, a constant size evolution with r$_{\rm e}$=3.6
kpc (the final average size of their sample). The results of our
model are in approximate agreement with these observations. Also in
Fig.~\ref{fig:SFIvsz} we show the stellar mass surface density evolution
of MW-like galaxies \citep{vandokkum13}.  It only evolves by about a
factor of 3 but includes the growth of the bulge so the evolution
in the effective mass surface density of the disk would be lower. A
small increase in the total stellar mass surface density is important
within the context of our model because there is also a dependence on
$\Sigma_{\star}$ and finding little or no evolution in the stellar mass
surface density \citep{barden05} suggests that the evolution in $\Sigma_{\rm
SFR}$, and hence the sSFR, is mostly due to the changing gas surface
densities and gas fractions.

To begin to understand why there is a change in the evolution of the sSFR
above and below z$\sim$2, we show the evolution of the star formation
intensity for galaxies above z$\sim$2 assuming the size evolution for
a galaxy with M$_{\star}$$\sim$5$\times$10$^9$ M$_{\sun}$ and sSFR=2
Gyr$^{-1}$ (approximately the mean sSFR of z$\ga$2 galaxies). The
generalized Schmidt law appears to over predict the star formation
intensities, suggesting some other regulatory mechanism comes into
play. Moreover, the star formation intensities are above the threshold
determined for starburst galaxies in the local universe to drive winds
\citep{lehnert96, heckman03} but this threshold may be somewhat higher
for high redshift galaxies \citep[1 M$_{\sun}$ yr$^{-1}$ kpc$^{-2}$;
][]{LT11b, newman12a}. This suggests that the regulatory mechanism is
related to the intensity of star formation (we discuss this subsequently
in Sec.~\ref{sec:regulatedSF}).

We note that such an hypothesis does not preclude observing strong
outflows from galaxies at z$\la$1-2 \citep[as they are observed,
e.g. ][]{lehnert96,rubin10, coil11, martin12, bouche12,rubin13}. Quite
the contrary. It just implies that the regions of high star formation
intensity are compact and occur only in the circum-nuclear regions
and that most galaxies do not drive winds or do so only briefly
\citep[e.g.][]{lehnert96}.

\subsubsection{The importance of mass return}

We should be cognizant that even though the gas supply from accretion
is likely declining, the mass return from the stellar populations
within galaxies can be significant and will be especially important
for (rotationally supported) galaxies with older stellar populations
\citep[$\sim$20\% or more of their total stellar mass,][]{leitner11,
snaith14}. This gas will have an angular momentum content similar to the
pre-existing disk, help to maintain the high angular momentum of the
accreted gas and aid in growing the disk \citep{martig10}. Of course,
objections are often raised about the relative contribution of mass
return compared to gas accretion, for example, the G-dwarf problem and
the abundance of deuterium in the local ISM. However, these problems may
not require gas accretion or at least not at significant rates \citep[see
e.g.][for a detailed discussion of these points]{haywood01, prodanovic10,
chaippini02, romano06, lagarde12, leitner11, snaith14}.

At z=0, the specific cosmological gas accretion rate is likely
significantly below the sSFR for actively star forming galaxies
\citep[e.g.][] {weinmann11, leitner12}. If such galaxies were forming
stars at an approximately constant rate, as observed for the Milky Way
over the last 3 Gyrs and perhaps much longer \citep{hernandez00,snaith14}
and as perhaps implied by the constant gas depletion timescale for local
star forming galaxies \citep{leroy13}, then the fraction of gas that has
been returned is about 20-40\% depending on the initial mass function
of the age weighted stellar masses \citep{leitner11}. The fraction of
the returned gas that is in the molecular phase is likely less than 50\%
and the distribution of the gas is related to the extent of the stellar
disk \citep{bigiel12}. This emphasizes both the importance of the angular
momentum and mass return from the stellar population in determining
the gas mass surface densities, which in turn, from our analysis of the
generalized Schmidt law, controls the evolution of the specific star
formation rate.

\subsubsection{Why the break in the sSFR evolution at
z$\sim$2?}\label{breakzeq2}

While overall it is likely that the decreasing gas supply plays
a fundamental role in determining the z$\sim$2 transition redshift, since this
redshift, above which the growth of the ensemble of galaxies appears to
be self-regulated, and below which angular momentum begins to dictate the
decline in the specific stellar mass growth rate with redshift, occurs
at approximately the same moment as when the cosmological specific gas
accretion rate becomes less than the average sSFR.

However, this is not to imply that gas accretion is necessarily the
main driver of the evolution of the sSFR at z$\la$2 but only determines
approximately when the transition occurs. In such a picture,
galaxies at z$\la$2 are living off of their gas-rich earlier phases of
evolution (and significant mass return) where star formation is kept
inefficient through strong feedback from intense star formation. The high
angular momentum of the gas that is being subsequently accreted allows
galaxies to preferentially grow their outer disks \citep{stewart11},
but this is likely moderated by the low stellar mass surface densities
of outer disks because as argued through the generalized Schmidt law,
their star formation should be kept relatively inefficient
\citep{blitz06}. Such a picture naturally explains the lack of a
significant increase in the central surface densities of star forming
galaxies from z$\sim$1 \citep{barden05, vandokkum13}.

It is not clear if the scenario we are advocating is consistent with
``inside-out'' evolution and the observation that the \HI\ mass surface
density does not depend strongly on radius in individual galaxies
\citep{bigiel12}. However, we may not expect the outer disk to grow more
rapidly than the inner disk even if the gas dominates the mass surface
density \citep{bigiel10, bigiel12} since interstellar pressure plays a
significant role in converting gas from the warm neutral phase to the
cold molecular phase \citep{wolfire95} and this conversion is related
to the stellar mass density \citep[e.g.][]{blitz06, schruba11} and of
course consistent with the generalized Schmidt law where the stellar mass
surface densities play an important role in regulating star formation
\citep[Sect.~\ref{sec:SFRmass} and also, ][]{shi11}.

The importance of this transition redshift and angular momentum is
emphasized by a number of observations of the stellar populations
of both the Milky Way and other nearby galaxies. Before the z$\sim$2
transition, the Milky Way, formed its thick disk, had high star formation
intensities and high stellar velocity dispersions, and its disk had a
short scale length \citep[e.g.][ and references therein]{haywood13}. The
low scatter in [$\alpha$/Fe] as a function of age and the rapid decrease
in [$\alpha$/Fe] with time of the thick disk suggests that the mixing of
metals was very efficient and the increase in metallicity was dominated
by core collapse supernovae. In addition, as the thick disk evolved,
its stellar velocity dispersion decreased \citep{haywood13}. The
young Milky Way formed during an intense burst of star formation with
increasing angular momentum which gradually died down over about 4
Gyrs and undoubtedly experience strong feedback and high turbulence
\citep{haywood13, snaith14, L14}. At around the transition redshift,
$\approx$10 Gyr ago, the abundance ratios and ages of stars suggest that
the MW starts to form stars in an outer thin disk while the inner disk
was still thick and as did perhaps the outer disks of other nearby spiral
galaxies \citep{ferguson01, vlajic09, barker11, haywood13}.  Overall, the
results on the Milky Way suggest an early phase of low angular momentum
gas accretion which fueled intense star formation with a short scale
length but large scale height followed by a phase of quiescent growth with
a longer scale length and a small scale height.  These are core arguments
in our scenario and appear to be mimicked in the star formation history
of the MW and other nearby galaxies. Such a picture is in agreement
with our analysis shown in Fig.~\ref{fig:SFIvsz}. At z$\ga$2, the star
formation intensity of galaxies is sufficiently high to drive strong winds
and create a highly turbulent ISM as observed \citep{newman12a, L13}.

Care must be taken in arguing for a continuous scenario of galaxy
evolution over the last $\sim$13 Gyrs. It is extremely likely that the
evolution of the sSFR of the ensemble of galaxies is driven by changing
populations at different redshifts. While the Milky Way shows evidence
for continued growth over the age of the universe \citep{haywood13,
vandokkum13}, massive galaxies, in particular, have stellar age
distributions dominated by old populations which formed relatively quickly
\citep[e.g.][]{johansson12} and have high stellar surface densities
\citep[e.g.][]{shen03, bernardi10, carollo13} consistent with them growing
primarily through high intensity star formation with strong feedback.

In addition, using the evolution of the sSFR itself to constrain the
growth of mass in the ensemble of galaxies also suggests that there
is a possible dichotomy in the growth rates of galaxies as a function
of mass, in that more massive galaxies grew rapidly above z$\sim$2
\citep{leitner12}. The high intensity star formation we observe is
consistent with this dichotomy but within the context of our strong
feedback scenario would also predict that the duty cycle of star formation
is relatively small ($\sim$10-20\%) at high redshifts \citep{verma07,
davies12}. Strong feedback limiting the duration of the duty cycle
may be consistent with a flat or slowly increasing sSFR with redshift
\citep{wyithe14} as are the general arguments we have made here.
Given that our analysis indicates that feedback processes
might regulate the SFR in galaxies at high redshift, we now discuss two
possible mechanisms for doing so.

\subsection{Mechanical and radiative self-regulated star
formation}\label{sec:regulatedSF}

It is possible that star formation is limited by its own mechanical and
radiative energy output (self-regulation). On global scales, there are
(at least) two possible mechanisms whereby the energy injected by massive
stars could limit the sSFR. One balances the pressure (thrust) generated
by the thermalized hot plasma due to the combined action of stellar
winds and supernovae plus radiation pressure with the overall galactic
mid-plane hydrostatic pressure. In the other, the high turbulence in
the dense molecular medium is sustained by a mass and energy exchange
within the ISM driven by the energy output of the young stars, which
maintains distant galaxies near or on the line of instability (i.e. Toomre
Q$\sim$1). It is this mass and energy exchange that ultimately regulates
the sSFR of distant galaxies. We will discuss these two in turn.

\subsubsection{Wind and radiative thrust balancing 
hydrostatic pressure}\label{subsec:windthrust}

We hypothesize that the mechanical energy from massive stars is
controlling the overall pressure, and in such a situation, we would
expect the pressure to increase linearly with the star-formation
intensity. The ultimate limit in the pressure driven by the mass and
energy output of massive stars is reached when it balances or exceeds the
mid-plane pressure -- similar to what has been hypothesized to limit the
star-formation intensity in nearby starburst galaxies \citep{lehnert96}.

The pressure due to the mechanical energy of intense star formation
is P$_{\rm feedback}$$\propto$$\dot{\rm M}^{1/2} \dot{\rm E}^{1/2}$
R$_{\star}^{-2}$, where P$_{\rm feedback}$ is the gas thermal
pressure generated by the effects of massive stars, $\dot{\rm M}$
is the mass loss, $\dot{\rm E}$ is the mechanical energy output and
R$_{\star}$ is the radius over which the energy and mass output occurs
\citep[e.g.][]{strickland09}. The constant of proportionality depends
on the opening angle (or geometry) of the flow, the thermalization
efficiency of the mechanical energy output, the mass entrainment rate in
the wind, and how well this energy and mass couples to the surrounding
ISM. Constraining these dependencies is difficult. From observations
of nearby starburst galaxies, the opening angle is approximately,
$\pi$ \citep{heckman90, lehnert96}. The thermalization efficiency
is likely to be high, about 0.3-1.0 \citep{strickland09} and the
mass loading (ratio of total outflow mass and total stellar mass
ejected through stellar winds and supernovae) high as well, about 4-20
\citep[e.g.,][]{moran99,bouche12}. In high redshift galaxies, the mass
loading may be higher than generally observed in nearby starbursts owing to
higher gas column densities and larger disk thicknesses \citep{lagos13}.

We can estimate the mechanical energy and mass output rate
from star formation using stellar population synthesis models
\citep{leitherer99}. Adopting an equilibrium mass and energy output rate
for continuous star formation over 10$^8$ yrs, we estimate pressures
of 3.6$\times$10$^{-10}$ dyne cm$^{-2}$ for 1 M$_{\sun}$ yr$^{-1}$
kpc$^{-2}$. Another estimate can be made from previous studies which found
that the outflow rate is roughly equal to the SFR. This
implies a mass loading/entrainment factor about 4 or a pressure of
7.2$\times$10$^{-10}$ dyne cm$^{-2}$ \citep[][]{moran99,strickland09}
but as noted previously, it could be higher as it scales with the square
root of the mass loading factor.

The hydrostatic pressure depends on the gas and stellar mass surface
densities (which are related through the gas fraction) and adopting
the more general relation, on the ratio of the gas to stellar velocity
dispersions \citep[\S~\ref{subsubsec:dispersions}; ][]{elmegreen93}.
We therefore need to estimate mass surface densities, gas fractions,
star formation rates and galaxy sizes.

Measurements of stellar mass surface densities in star-forming
galaxies at z$\approx$1-6 estimate $\sim$100-2000 M$_{\sun}$ pc$^{-2}$
\citep[e.g.][]{barden05,yuma11, fs11, mosleh12} and high gas fractions
\citep[e.g.][]{tacconi13}. At z$\sim$5-7, galaxies are estimated to have
a very high sSFR, $\approx$10 Gyr$^{-1}$ \citep{stark13}, although the
uncertainties in this value are large. The observed fiducial stellar mass
of these galaxies is $\sim$10$^9$ M$_{\sun}$ and their half-light radius
$\sim$0.5 kpc \citep{mosleh12}. These are intensely star forming galaxies,
with $\Sigma_{\rm SFR}$$\approx$6 M$_{\sun}$ yr$^{-1}$ kpc$^{-2}$ and high
stellar mass surface densities, $\Sigma_{\star}$$\approx$600 M$_{\sun}$
pc$^{-2}$. The more massive galaxies with the smallest half-light radii at
these redshifts can reach $\Sigma_{\star}$$\approx$1000-2000 M$_{\sun}$
pc$^{-2}$, similar to early type galaxies at moderate to low redshifts
\citep{carollo13}.

At lower redshifts, z$\approx$2, the average mass surface density is
much lower, thus making it easier in this scenario for the mechanical
energy to regulate the star formation through pressure. In galaxies at
z$\approx$2-3, $\Sigma_{\star}$$\approx$200-300 M$_{\sun}$ pc$^{-2}$
\citep{fs11, mosleh12} and their average $\Sigma_{\rm SFR}$$\approx$1
M$_{\sun}$ yr$^{-1}$ kpc$^{-2}$ \citep{L13}. The gas fractions of galaxies
at z$\sim$5-7 are unknown, but at z$\sim$2, they are observed to be
f$_{\rm g}$=30-70\% \citep{tacconi10,tacconi13}. As already discussed
(\S~\ref{sec:SFRmass}), we do not know the velocity dispersions of the gas
relative to the stars and their ratios could easily range from 0.2 to 1.

\begin{figure}
\includegraphics[width=8.95cm]{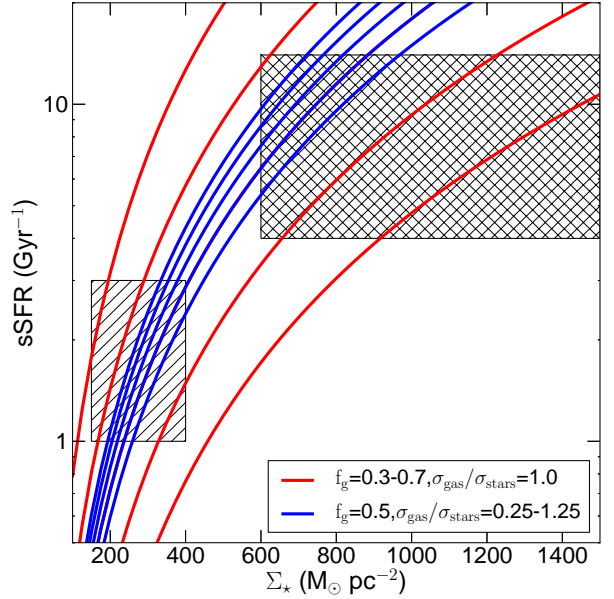}
\caption{The specific star formation rate (sSFR, in Gyr$^{-1}$) as
a function of the stellar mass surface density ($\Sigma_{\star}$,
in M$_{\sun}$ pc$^{-2}$) as predicted by a simple relation where the
hydrostatic pressure and the thrust of the mechanical energy generated
by intense star formation are equal. To span a range of possible sSFR
in such a scenario, we vary the input values of the gas fraction,
f$_{\rm g}$, or the velocity dispersion ratio of the gas and the stars,
$\sigma_{\rm gas}/\sigma_{\rm stars}$. The red solid lines represent this
equality for a range of gas fractions, f$_{\rm g}$=0.3 to 0.7, and for a
constant $\sigma_{\rm gas}/\sigma_{\rm stars}$=1, while the blue lines
represent this equality for a constant gas fraction, f$_{\rm g}$=0.3,
and a range of $\sigma_{\rm gas}/\sigma_{\rm stars}$=0.25-1.25 in steps
of 0.25. The hatched regions indicate approximately the observed range of
stellar mass surface densities and sSFR at z$\sim$2 (diagonally hatched
region) and z$\sim$5-7 (cross-hatched region; see text for details). }
\label{fig:sSFRevolmodels}
\end{figure}

When we simply equate the hydrostatic pressure
(\S~\ref{subsubsec:dispersions}) and the ``feedback pressure'', we find
reasonable agreement with observed stellar mass surface densities and sSFR
for galaxies at z$\sim$2 and z$\sim$5-7 (Fig.~\ref{fig:sSFRevolmodels}),
for a reasonable range of f$_{\rm g}$ and $\sigma_{\rm gas}$/$\sigma_{\rm
stars}$, assuming a thermalization efficiency of 0.5, a mass loading
factor of 10 or an outflow rate of 2-3 times the SFR.

We emphasize that within the context of this simple calculation,
the mass loading factor and thermalization efficiency are degenerate
and that it is the product of the square root of their values that
results in the estimate of the mechanical energy. Since the half-light
radius of galaxies at constant stellar mass evolves systematically with
redshift, such a model provides a reasonable explanation for the sSFR
at z$\approx$2-7. This simple scenario predicts a strong decline in
the sSFR with stellar mass surface density and some of the curves in
Fig.~\ref{fig:sSFRevolmodels} for high gas fractions and large ratios
of gas to stellar velocity dispersions lie above the region occupied by
distant galaxies.

The equivalent of requiring the feedback pressure to balance or exceed
the hydrostatic pressure is to say the gas is only marginally bound.
Obviously, this is an extreme assumption and unlikely to hold true over
the entire ISM simultaneously. Therefore, it is perhaps more appropriate
to consider this as a viable mechanism for limiting the values that the
SFR can sustain.

Using the mechanical energy output from young stars to limit the sSFR is
purely empirical in that it does not make predictions of the evolution of
the sSFR but only provides upper limits or ranges for it, which depend
on parameters that can solely be determined observationally. In some
sense, it is similar to phenomenological models of galaxy evolution,
though with significantly less complication, in that it attempts
 physical descriptions of processes that are otherwise
difficult to constrain, such as the star formation efficiency or the
relation between star formation intensity and gas surface density
\citep[e.g.][]{behroozi13,feldmann13}. While we have not tried to
constrain our input parameters by fitting the evolution of the sSFR,
and rather adopted values consistent with observations, it indicates that
this type of scenario works reasonably well in describing the evolution
of the sSFR.

Since the mechanical energy input depends on the mass loading and the
thermalization efficiency, it is quite likely that as the galaxies
grow less compact, the area covered by the intense star formation
decreases and the overall pressure of the ISM will drop in lock step
with declining hydrostatic pressure. However, galaxies at z=2-4 exhibit
clumpy star formation with ample evidence for driving significant outflows
\citep{LT11b,newman12a,newman12b} and the thermal pressure of the warm
ionized medium in these clumpy galaxies appears high \citep[similar
to nearby starburst galaxies driving winds;][]{L09, L13, LT11} and
perhaps also have high turbulent pressures in the cold molecular medium
\citep{renaud12, L13}. This clumpy structure likely leads efficient
thermalization and coupling of the mechanical energy output of young
stars to the surrounding ISM, allowing the mechanical (and radiative)
energy to influence the gravitational collapse of dense gas and therefore
the SFR and the sSFR. These individual clumps of intense star formation
are akin to individual galaxies at z$\approx$6, in the sense that they are
similarly compact and that the most massive ones can have similar stellar
masses \citep{elmegreen09,elmegreen09a,fs11} and sSFR \citep{guo12},
and they show evidence for strong stellar feedback \citep[e.g.][]{LT11b,
newman12a}. These similarities and relationships may allow for similar
regulation of star formation and baryonic growth but less efficiently in
the case where the most intense star formation covers the entire
disk, because of the lower covering factor of similarly intense star
formation within the (significantly larger) disk \citep[][]{fs11, guo12}.

Certainly, with reasonable parameters, finding a decrease of about a
factor of 3 in the sSFR with stellar mass surface density is consistent
with the overall trend observed (Fig.~\ref{fig:sSFRevolmodels}). The
normalization appears high compared to the data, but it is important to
note that our underlying assumption of an almost unbound disk is quite
extreme. We would expect such a model to explain the most intensely star
forming galaxies, perhaps not the typical ones.

%fig 3

\subsubsection{Turbulent pressure and the line of stability
(Q$\sim$1)}\label{subsubsec:Q1} 

Galaxies at high redshift, z$\ga$1, have broad optical emission lines
\citep[velocity dispersions $\sim$30-200 km s$^{-1}$; e.g.,][]{Epinat09,
FS09, law09, epinat10, kassin12} which may be related to their high
observed star-formation intensities \citep{L09, L13}. In \citet{L13},
we proposed that there is a mass and energy exchange in the ISM of
distant galaxies between the warm ionized medium and the cold molecular
medium. The high pressures observed in the warm ionized medium (and
its likely similarity with the hot X-ray emitting gas) implies that
much of the ionized gas will quickly become cold neutral gas and, if
it contains dust, cold molecular gas \citep{wolfire95,feldmann12}. This
rapid phase change under high pressures allows the cold molecular gas to
``capture'' the kinematics of the warm ionized gas. In such a scenario,
the molecular gas acquires much of the turbulent motions observed in the
optical emission line gas and since the molecular phase dominates the
overall mass surface density, it may have sufficient turbulent pressure to
approximately balance gravity over large scales. This balance then leads
to a situation where galaxies are driven towards the line of stability for
global star formation (i.e., a Toomre instability criterion of Q$\sim$1).

In such a picture, if the star formation intensity were to increase, the
galaxy would move beyond the line of stability and star formation would
be suppressed; if the star formation intensity falls, the galaxy would
become globally more unstable and thereby increase its star formation
intensity. This naturally leads to the regulation of star formation to
a narrow range around the line of instability, Q$\sim$1.

A sufficiently high gas content is needed to regulate star formation.
Star formation, as observed in local galaxies, becomes regulated by
the local balance of turbulent energy dissipation, energy released by
gravitational collapse and instability, and energy injected by the
stellar population. Such balance would no longer be maintained when
the gas content decreases to the point when it is either insufficient
to fuel the necessary high star formation intensities, when the outflow
of gas and energy are significant enough to make the energy input from
massive stars unable to sustain high levels of disk turbulence, or when
the turbulent dissipation timescale becomes long enough such that star
formation becomes inefficient (efficient dissipation is necessary to
sustain intense star formation). The disk would then settle over time
as the gas mass surface densities decrease \citep[perhaps as observed,
e.g.][]{epinat10, kassin12}.

If we assume that galaxies lie near the line of instability and that
the turbulence in the ISM is driven by intense star formation of the
form, $\sigma_{\rm gas}$=$\epsilon\Sigma^{1/2}$ (where $\epsilon$ is
the coupling factor between the star formation intensity, in units of 1
M$_{\sun}$ yr$^{-1}$ kpc$^{-2}$, and $\sigma_{\rm gas}$ is the velocity
dispersion of the gas in km s$^{-1}$) then, applying the effective Toomre
criterion \citep[${{\rm Q}_{\rm total}}^{-1}$=${{\rm Q}_{\rm stars}}^{-1}$
+ ${{\rm Q}_{\rm gas}}^{-1}$;][]{wang94}, we find,

\begin{equation}
\nonumber
{\rm sSFR} = \left(\frac{\Sigma_{\rm SFR}}{\Sigma_{\star}}\right) =
\left(\frac{\pi {\rm G} {\rm Q}_{\rm total}}{\kappa \epsilon}\right)^{2} \Sigma_{\star}\left(\frac{{\rm f}_{\rm g}}{1-{\rm f}_{\rm g}} + \frac{\sigma_{\rm gas}}{\sigma_\star}\right)^{2}
\end{equation}

where $\kappa$ is the epicyclic frequency (taken to be $\sqrt{2}$
$\Omega$, with $\Omega$ the angular frequency, and assuming
a constant circular velocity) and f$_{\rm g}$ is the gas fraction.

The derivation assumes that the total sSFR=SFR/M$_{\star}$ = $\Sigma_{\rm
SFR}/\Sigma_{\star}$. In \citet{L09} and \citet{L13}, we estimated
that an efficiency factor of 140 km s$^{-1}$ (M$_{\sun}$ yr$^{-1}$
kpc$^{-2}$)$^{-1/2}$ was necessary to explain the relation between
the star formation intensity and the velocity dispersion in galaxies
at z$\approx$2 \citep{L13}. However, this factor could be 30\%
lower, as we generally did not correct for extinction in the sample,
and the star formation intensities could be higher by a factor of about
2 \citep{L09, L14}. In Fig.~\ref{fig:sSFRevol}, we assumed 120 km
s$^{-1}$ (M$_{\sun}$ yr$^{-1}$ kpc$^{-2}$)$^{-1/2}$ to take into account
that the efficiency is likely lower \citep{L14}.

We note that this analysis is closely related to that given in
\S~\ref{sec:SFRmass} based on the generalized Schmidt law, which we used
to estimate the relationship between the SFR and stellar
mass (Fig.~\ref{fig:elbaz}) since both arguments are ultimately related
to the pressure in the ISM.

\begin{figure}
\includegraphics[width=8.95cm]{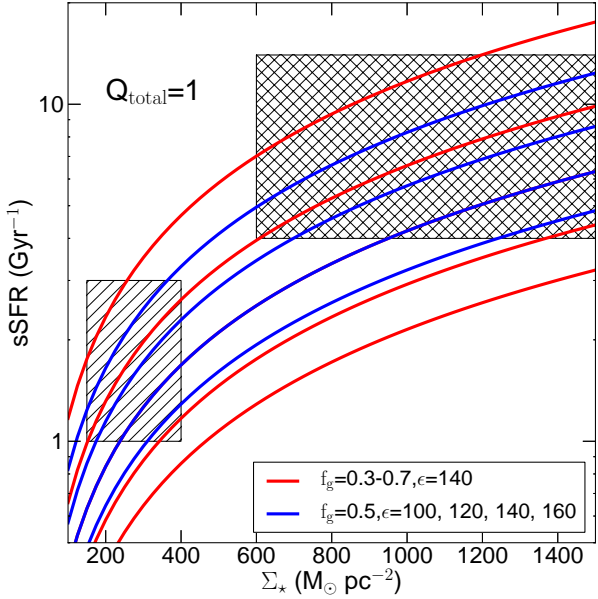}
\caption{The specific star formation rate (sSFR, in Gyr$^{-1}$) as a
function of stellar mass surface density ($\Sigma_{\star}$, in M$_{\sun}$ pc$^{-2}$), for the hypothesis that the star formation drives galaxies towards the line of stability against star formation (Toomre parameter, Q=1) as described in the text. The red and blue solid lines represents the sSFR where the energy input from massive stars generates enough turbulence to drive the global ISM towards Q$\sim$1. The red solid lines represent where Q$\sim$1 for a range of gas fractions (f$_{\rm g}$=0.3 to 0.7) and for a constant $\sigma_{\rm gas}/\sigma_{\rm stars}$=1, while the blue lines represent where Q=1 for a constant gas fraction (f$_{\rm g}$=0.5) and $\sigma_{\rm gas}/\sigma_{\rm stars}$=1, but now for a range of energy input from massive stars into the turbulence of the ISM, $\epsilon$ = 100--160 km s$^{-1}$ kpc yr$^{1/2}$ M$_{\sun}^{-1/2}$, where $\epsilon$ is the coupling between the star formation intensity and the velocity dispersion \citep{L09, L13}. The predicted sSFR values increase with increasing gas fraction but decrease with increasing $\epsilon$. The hatched regions are the same as in Fig.~\ref{fig:sSFRevolmodels}.}
\label{fig:sSFRevolmodels2} 
\end{figure}

The resulting relation shows a reasonable agreement with the observed
values for both sSFR and stellar mass surface density of distant galaxies
(Fig.~\ref{fig:sSFRevolmodels2}) and for the evolution of the sSFR with
redshift (Fig.~\ref{fig:sSFRevol}). For the latter, we scaled the sizes of
the galaxies by (1+z)$^{-1.2}$ for a constant stellar mass of 10$^{9.3}$
M$_{\sun}$, a gas fraction of 0.5, a ratio of gas to stellar velocity
dispersion of 1 and assumed a constant epicyclic frequency consistent
with those estimated at z$\approx$2-3 \citep{L13}, as we expect it to
only vary by a factor of a few with redshift.

Since we have some freedom of choice in the parameters we adopt, the
agreement may seem rather fortuitous. On the other hand, both models,
one in which supernovae and strong stellar winds determine the pressure
and balances hydrostatic equilibrium, and the other, where star formation
drives strong turbulence holding galaxies near the line of instability,
provide reasonable evolution in the sSFR with stellar mass surface density and redshift (although the thrust balancing the hydrostatic pressure is a very extreme assumption). The results of both models emphasize the important role that self-regulation of star formation plays in balancing the relative rate of growth of galaxies and the high gas fractions that are necessary to allow this stellar feedback to couple efficiently to the ISM \citep{L13}.

%fig 4 

\subsubsection{Comparison with other sSFR evolution models}

Although there have been other studies of the possible underlying
processes which may dictate the sSFR-M$_{\star}$ relation and
its evolution over cosmic time, there is as yet no consensus on
what these may be. Only a few studies have suggested that self
regulation through interstellar gas pressure is the primary driver
of the observed evolution \citep[e.g.][]{birnboim13}. The study of
\citet{dutton10} finds that the evolution of the sSFR is driven by
the details of the gas accretion history, and by increasing both
the gas surface densities of molecular gas and the ratio of the
molecular to atomic gas mass \citep{blitz06}. Surprisingly, they do
not predict an evolution of the gas fraction, except at the highest
redshifts. Other models suggest that the evolution at z$\la$2--4 can
be entirely explained by changes in the specific gas accretion rate
\citep[e.g.][]{dutton10,kang10}. Interestingly, it is over this redshift
range that there is almost no evolution in the mean density of halos
at constant halo mass within 20 kpc radius despite the large growth in
the virial radius \citep{weinmann13}. At z$\ga$4, \citet{khochfar11}
suggest that the sSFR is driven by the star-formation efficiency
being dependent on the mode of accretion -- mergers or cosmological
gas accretion. Their model predicts a preponderance of mergers at high
redshift. \citet{weinmann11}, in a comprehensive study, found a number
of ways in which a plateau, or a slowly rising sSFR at z$\ga$2 may be
explicable. The effects they suggested include a reduced star-formation
efficiency/enhanced feedback, prohibiting the gas from forming stars,
efficiently ejecting the gas which is subsequently accreted by more
massive halos, or through an enhanced growth rate of massive galaxies.

The advantage of our simple scenario, in comparison, is that the gas
accretion history has little impact on the evolution of the sSFR with
cosmic time, only that sufficient gas needs to be accreted at high
redshifts to allow the gas content to build up high mass surface density
galaxies at high redshift, that the angular momentum of the accreted
gas generally increases with decreasing redshift, and that the rate
of gas accretion then declines with decreasing redshift. The evolution
of the sSFR in our model is determined by the change in the angular momentum 
content of the accreted gas \citep{dubois13,danovich14} and how the gas and
stellar mass surface densities are limited by the energy and mass output
of massive stars. With these two processes, the range of gas/stellar
mass surface densities observed in galaxies as a function of redshift
may be explained. The star formation intensities are related to the gas
mass surface density through the well-known simple relation between the
two with an exponent of the order of unity \citep{leroy13}. Higher gas mass
surface densities lead to higher star-formation intensities, which
in turn lead to higher stellar mass surface densities, both of which
are limited by feedback from massive stars but whose overall evolution
is driven by the angular momentum with which gas is accreted. This is
naturally proportional to both the gas fraction, total gas content and
total mass of galaxies. It requires little further fine tuning in that
at high redshift (z$\ga$2) the gas supply is in excess of that needed
to support star formation and that at low redshift (z$\la$2) centrifugal
support is important in limiting the ultimate surface density (or volume
density) of gas in galaxies. Importantly, it is this natural limiting of
the surface densities through the pressure that gives the SFR-M$_{\star}$
relation its slope of one -- consistent within the uncertainties with
studies of the SFR-M$_{\star}$ relation.

\section{Summary}\label{summary}

We find that the SFR-M$_{\star}$ relationship and the evolution of
specific star formation rates with cosmic time are consistent with
a scenario where the relative growth rates of galaxies are set by
the interplay of the stellar and gas mass surface densities and by
self-regulated star formation. The stellar mass surface densities are
related to the gas mass surface densities through the relationship between
star-formation intensity and gas mass surface density, whose exponent
is approximately unity \citep{leroy13}. It is the angular momentum of the
accreted gas that will help set the ultimate limit on the intensity of the
star formation by influencing the surface densities the gas is able to
reach. Feedback and self-regulation from young stars help keep the star
formation inefficient. The slope of the SFR-M$_{\star}$ relationship,
which is a ridge line in the SFR-M$_{\star}$ plane, can successfully
be reproduced in a scenario where the global SFR in
galaxies is related to the overall pressure in the ISM which is itself
due to the intense star formation \citep{hopkins13}. This also implies
that, even when the gas supply through cosmic accretion is very large,
the sSFR cannot increase without bound, but only slowly, with limits set
by feedback from intense star formation within the context of the high
stellar mass surface densities observed in the early universe (z$\ga$2-3).

The decrease in sSFR at z$\la$2 implies that the gas mass surface density
and gas supply are reduced below the levels necessary to maintain the high
intensity, compact star formation as observed. The growth of the ensemble
of galaxies is no longer self-regulated by star formation. Now, with the
gas supply below this threshold, processes internal to the galaxy or the
mode of accretion become important. In a simple cosmological model, the
radius of a centrifugally supported gaseous disk is expected to evolve
as (1+z)$^{-1.5}$ and thus the mass surface density (at constant mass)
as (1+z)$^{3}$/t$_{\rm H0}$ -- similar to the observed rate of decline
of the sSFR with redshift at z$\la$2 \citep{oliver10, elbaz11}. Thus we
hypothesize that what limits the star formation efficiency and baryon
content of galaxies to the level necessary to explain the observed
evolution of the sSFR with cosmic time is a combination of the gas
accretion rate (whether it is generally higher or lower than the sSFR
which then established the transition redshift between the two regimes
controlling the sSFR), the angular momentum of the accreted gas (which
then determines both how compact the galaxy is at a constant stellar mass
and the evolution of its gas fraction), and self-regulating star formation
and feedback through regulating the pressure of the ISM. Obviously, more
work is needed beyond these simple ideas to gauge if the scenario we
have advocated is consistent with other constraints on galaxy evolution
(e.g. cosmic star-formation history, size evolution of galaxies, etc.).

\begin{acknowledgements} 

MDL wishes to thank Johan Dubois, Gary Mamon, S\'ebastian Peirani,
and Marta Volonteri for interesting discussions, and Christophe Pichon
and Mike Fall for the engaging talks at the IAP 75$^{th}$ anniversary
conference which helped some of these ideas to congeal. We especially
thank David Elbaz, Gary Mamon, and Joe Silk for their critical reading
of and helpful comments on this manuscript.  We thank our former
co-author N. P. H. Nesvadba for her participation in the preparation of
this manuscript.

\end{acknowledgements}

\bibstyle{aa}
\bibliographystyle{aa}

\bibliography{sSFRrefv1}

\begin{appendix}
\section{On equation~\ref{eqn:SFR}}

The derivation of equation~\ref{eqn:SFR}, which is central to our
analysis, is given in \citep{silk09} but we will repeat some of their
arguments here for completeness. The generalized Schmidt law of this form
can be derived through a cloud-cloud collision model where the velocity
dispersion between clouds is driven by the energy input from supernovae
\citep{silk09}. Since the dispersion in the clouds is a pressure, ${\rm
P}_{\rm gas}$=$\rho_{\rm gas}\sigma_{\rm gas}^2$, and in equilibrium
it can be related to the hydrostatic pressure, ${\rm P}_{\rm gas}$=
$\frac{\pi}{2} {\rm G}\Sigma_{\rm gas}\Sigma_{\rm total}$.  Thus within
the context of this model, the star formation intensity is related to
the pressure in the ISM.

The normalization of the generalized Schmidt law (see
Sect.~\ref{subsec:simplemodel}) deserves some brief discussion. As shown
in \citet{silk09} (their equation 4):

\begin{equation}
\Sigma_{\rm SFR} = f_{c} f_{cl} G (\pi \Sigma_{\rm total})^{1/2} {{m_{\rm SN} v_{c}}\over{E_{\rm SN}}} ({{p_{g}}\over{p_{cl}}})^{1/2} \Sigma_{\rm gas}^{3/2}
\label{eqn:SFRder}
\end{equation}

where ${p_g}\over{p_{cl}}$ is the ratio of the gas pressure to the
internal cloud pressure, $f_c$ and $f_{cl}$ are the cloud filling
factor and the fraction of the cold neutral and molecular gas in clouds
respectively, and $m_{SN}$, $v_c$ and $E_{\rm SN}$ are the mass of stars
formed per SNe II (150 M$_{\sun}$ for a Chabrier IMF), the velocity at
the onset of strong cooling in the remnant and the kinetic energy of
SNe II, respectively.

To derive equation~\ref{eqn:SFR}, we multiplied the right hand side of
equation~\ref{eqn:SFRder} by $\Sigma_{\rm total}/\Sigma_{\rm total}$
(=1), and rearranged the terms to yield,

\begin{equation}
\Sigma_{\rm SFR} = f_c f_{cl} G \pi^{1/2} {\Sigma}_{\star}\ \frac{{\rm f}_g^{1/2}}{(1-{\rm f} _g)} \Sigma_{\rm gas}
\label{eqn:SFRder2}
\end{equation}

The total SN energy input into the ISM is 10$^{51}$ erg, for an
assumed canonical supernova energy transfer efficiency into momentum of
the ISM of 0.01, and $v_c$ = 400 km s$^{-1}$ \citep{silk09}.  The relevant
quantities for the distribution of the gas, $f_c$ and $f_{cl}$, are
not well constrained in galaxies, but reasonable numbers are $f_c$=0.2
and $f_{cl}$=0.1, and these clouds are mildly over pressurized relative
to the ISM (${p_g}\over{p_{cl}}$ is approximately a few; where we have
chosen 2).

To derive the final version of the relationship (equation~\ref{eqn:SFR}),
we simply multiplied equation~\ref{eqn:SFRder2} by the disk area (i.e.,
SFR=2$\pi$r$_{\rm e}^2$$\Sigma_{\rm SFR}$ and M$_{\star}$=2$\pi$ r$_{\rm
e}^2$$\Sigma_{\star}$). Thus we assumed that the area covered by star
formation is that of the stellar disk, which of course is the fundamental
assumption in a generalized Schmidt law.

\end{appendix}

\end{document}